\documentclass[epj,final]{svjour}

\usepackage{graphicx}
\usepackage{dcolumn}

\newcommand{\tbtbt}{$2$$\times$$2$$\times$$2$ }
\newcommand{\obobo}{$1$$\times$$1$$\times$$1$ }
\newcolumntype{d}{D{.}{.}{3}}

\begin{document}

\title{TiC Lattice Dynamics from ab initio Calculations}

\author{P.T. Jochym \and K. Parlinski \and M. Sternik}
\institute{Institute of Nuclear Physics, 
  ul. Radzikowskiego 152, 31--342 Cracow, Poland
  \mail{jochym@ifj.edu.pl}}

\abstract{ {\it Ab initio} calculations and a direct method have been
  applied to derive the phonon dispersion curves and phonon density of
  states for the TiC crystal. The results are compared and found to be
  in a good agreement with the experimental neutron scattering data.
  The force constants have been determined from the Hellmann-Feynman
  forces induced by atomic displacements in the \tbtbt supercell. The
  calculated phonon density of states suggests that vibrations of Ti
  atoms form acoustic branches, whereas the motion of C atoms is
  confined to optic branches. The elastic constants have been found
  using the deformation method and compared with the results obtained
  from acoustic phonon slopes.  }


\maketitle


The transition-metal carbide compounds, of which TiC is a
representative, are of big scientific and technological interest
because of their striking mechanical properties, extreme hardness
combined additionally with metallic electrical and thermal
conductivity.  They have usually a rock-salt crystal structure -- like
some ionic crystals -- while having properties typical for materials
with covalent bonding.  It suggests that the bonding have some mixed
nature, which makes these materials very interesting as an object of
study.
 
Over a large range of fractional carbon content from TiC to
TiC$_{0.5}$ \cite{toth} the TiC is stable in the NaCl structure, with
space group Fm${\bar 3}$m.  In practice, it is usually
nonstoichiometric and contains carbon vacancy defects.  The phonon
dispersion curves of nearly stoichiometric TiC$_{0.95}$ have been
measured along main symmetry directions by Pintschovius et al
\cite{pint} using the inelastic neutron scattering technique.  In the
same paper, the influence of the carbon content on the phonon
dispersion curves has been additionally checked by measuring a sample
with lower carbon concentration TiC$_{0.89}$.  The results shown that
the optic modes could differ by 3.5\%, while the acoustic modes are
less sensitive to carbon content.  The elastic constants derived from
the measured acoustic dispersion curves agree very well with those
obtained by ultrasonic measurements for TiC$_{0.91}$ crystal
\cite{exp_dataA}.

Very interesting properties and simple structure of TiC caused that
efforts to understand its properties using first-principle
calculations have been undertaken several times
\cite{neckel,zhukov,price,haglund}.  The electronic structure, bulk
modulus, and elastic constants have been found \cite{TiC_GGA} by means
of accurate first-principles total-energy calculations using the full
potential linear muffin-tin orbital method.  The calculated values are
generally in very good agreement with experiment.

In this paper we intend to extend first-principle calculations to
describe the phonon dispersion curves and phonon density of TiC.  The
method which was is based on the total energy calculation and {\it
  Hellmann--Feynman\/} (HF) forces.  Dispersion relations are
calculated by {\em direct method\/}
\cite{frank,kresse,parlinski,ackland,sluiter}, in which the force
constants of the dynamical matrix are calculated from HF forces.
Elastic constants and bulk modulus have been estimated by straight
evaluation of energy derivatives with respect to deformation.

The energies and forces of the TiC crystal were calculated by the
method of total energy minimization, using a norm--conserving {\em
  pseudopotentials\/} as an approximation of the atomic core --
valence electron interaction \cite{lin,goniakowskiA}. For an excellent
review of {\it ab initio\/} total--energy calculations see
Ref.\cite{payne} and further references given there. For the lattice
dynamics calculations the \tbtbt supercell with periodic boundary
conditions and 64 atoms has been used. For optimizing the structure
and for direct calculation of energy deformations \obobo supercell has
been utilized. The {\it ab initio} total energy calculations were done
with the CASTEP package \cite{castep} and standard pseudopotentials
provided within this package. These pseudopotentials were constructed
within LDA approximation. The Ti pseudopotential treats 3p electrons
as belonging to the core. We have employed non--local variant of those
potentials parametrized in the reciprocal space with 900 eV and 600 eV
cut--off energy for \obobo and \tbtbt supercell, respectively. Tests
which had been made with the cut-off energy of 600 eV applied to the
\obobo supercell, showed that in this case the cohesive energy is only
0.2 eV higher, and the lattice constant is longer by 0.0006\AA
(0.01\%), comparing to the 900 eV cut-off.  Therefore, we have used
600 eV cut-off energy for a larger supercell calculation.

A {\em generalized gradient approximation\/} (GGA) has been used for
the exchange energy term of the valence states of the Hamiltonian
\cite{perdew}. Although this procedure is not quite consistent, there
are indications \cite{goniakowskiB} that LDA generated
pseudopotentials used with GGA exchange term in crystals are
sufficiently accurate for energies and structures. The integration
over the Brillouin zone has been performed with weighted summation
over wave vectors generated by Monkhorst-Pack scheme \cite{pack} using
0.1 \AA$^{-1}$ and 0.06 \AA$^{-1}$ grid sizes which leads to 4 and 32
wave vectors for \obobo supercell and to 1 and 4 wave vectors for
\tbtbt supercell, respectively. No significant difference due to
different number of wave vectors, 1 or 4, has been found in calculated
phonon frequencies. The results of dispersion curve presented here are
generated from the 0.06 \AA$^{-1}$ data set with 4 wave vectors.

\begin{table*}[htbp]
  \begin{center}
    \begin{tabular}{|c|d|D{.}{.}{-1}|dddd|}
      \hline
      Quantity & \multicolumn{1}{c|}{Exper.} & 
      \multicolumn{1}{c|}{Non-metallic} & &
      \multicolumn{2}{c}{Metallic} & \\
      \hline
      \# k & & 
      \multicolumn{1}{c|}{4} & 
      \multicolumn{1}{c}{4} & 
      \multicolumn{1}{c}{14} & 
      \multicolumn{1}{c}{32} & 
      \multicolumn{1}{c|}{63} \\
      \hline
      $a_0$ & 4.327 & 4.345  & 4.35 & 4.342 & 4.345 & 4.342 \\
      $B$   & 2.4    & 2.42   & 2.41 & 3.527 & 2.418 & 2.427 \\
      $\omega_{\Gamma}$ & 16.3 & 16.5 & 19.0 & 11.8 & 15.6 & 15.6 \\
      $\omega_X^1$      &  8.5 &  8.5 & 10.2 &  5.1 &  9.2 &  8.5 \\      
      $\omega_X^2$      & 10.8 & 10.0 & 11.1 & 10.3 & 10.7 & 10.7 \\
      $\omega_X^3$      & 16.5 & 17.0 & 18.2 & 12.8 & 15.7 & 15.3 \\
      $\omega_X^4$      & 18.5 & 17.8 & 19.3 & 16.7 & 18.1 & 17.7 \\
      \hline
    \end{tabular}
    \caption{Comparison of results from metallic and non-metallic approach}
    \label{tab:metalic}
  \end{center}
\end{table*}

Since TiC is a weakly metallic compound it could have been appropriate
to use a smearing of electron levels in the Brillouin zone
integration.  This procedure is computationally more expensive. Thus
we have performed a series of tests to compare lattice constants, bulk
moduli and phonons in $\Gamma$ and X points obtained using gaussian
smearing in the range from 4 to 0.1 eV and denser k-space grids (4,
14, 32, 63 k-points) for one unit cell configuration. The comparison
of results shown in the Table \ref{tab:metalic} indicate that, for
sufficiently dense k-space grid used, these quantities are not very
sensitive (specially the lattice constant and bulk modulus) to the
choice of method of Brillouin zone integration. Thus we have decided
to use much simpler and faster non-metallic approach in the final
\tbtbt supercell calculation. Note also that results obtained in the
metallic regime lead to larger divergency of phonon frequencies from
experimental data.

The minimization of the total energy with respect to the lattice
constant was performed within the CASTEP minimizer module with \obobo
supercell and cross-checked with \tbtbt supercell.  The equilibrium
lattice constant is $a=4.3448$\AA{} which could be compared with the
experimental value $a=4.3269$\AA \cite{naray}. Such a good agreament
may be coincidental, since inclusion of 3p electrons in the valence
band as well as gradient corrections in core states tends to
change the equilibrium lattice constant \cite{garcia,juan,seifert}.

The HF forces are defined as
\begin{equation}
  {\bf F}_i({\bf n}, \nu ) = 
  - \partial E_{tot}/\partial {\bf R}_i({\bf n}, \nu)
  \label{HF}
\end{equation}
where ${\bf n}, \nu $ are the indices of the unit cell and atom within
the unit cell, respectively, and $i$ is the Cartesian component.  At
extremum all HF forces vanish.  Non-zeroth HF forces arise, when a
single atom $({\bf m}, \mu )$ is displaced by $u_j({\bf m}, \mu)$ from
its equilibrium position.  The arising forces are related with the
cummulant force constants $\Phi_{ij}$ by a relation
\cite{parlinski,ackland}
\begin{equation}
  {\bf F}_i({\bf n}, \nu ) = 
  - \sum _{{\bf m}, \mu ,j}
  \Phi _{ij}({\bf n}, \nu ; {\bf m}, \mu )
  u_j({\bf m}, \mu )
  \label{force}
\end{equation}
Cummulant force constants appear as a result of periodic boundary
conditions imposed on the supercell.  To calculate HF forces an atomic
configuration with a single displaced atom must be minimized with
respect to the electronic part only.  Each of such runs provides $3n$
HF forces, where $n = 8$ or $64$ is the number of atoms in the \obobo
or \tbtbt supercell, respectively.  Two runs with displacements along
$z$ direction, one for Ti and second for C were performed.  To
minimize the systematic errors two other runs with negative
displacements were carried on.  The displacement amplitudes were set
to $1\%$ of the lattice constant. This amplitude has been selected
after number of careful tests in which displacements ware varied from
$0.1\%$ to $1.6\%$. It has been proven that the anharmonic
contributions are negligable up to $1\%$ of displacement. At smaller
displacements, close to $0.1\%$ the HF forces became too small and the
uncertainty of force constant evaluation is not acceptable.

\begin{figure}[htbp]
  \leavevmode
  \includegraphics[width=\columnwidth]{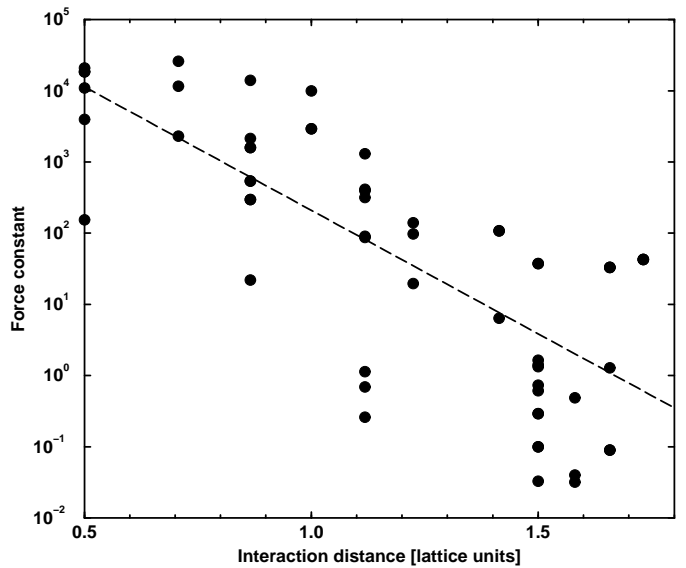}  
  \caption{Absolute eigenvalues of the
    force constant matrices of the TiC crystal as a function
    of interatomic distance.}
    \label{fig:force_const}
\end{figure}

The data of $4\times 3n$ HF forces ${\bf F}_i({\bf n}, \nu )$ and four
displacements $u_j({\bf m}, \mu )$ form an overdetermined set of
equations, Eq.(\ref{force}), for the force constants.  This system was
solved by the singular value decomposition algorithm
\cite{parlinski,numrec}, which automatically provides the
least-squares solution.  For \tbtbt supercell the 768 components of
the HF forces lead to 33 non-zero independent parameters of the force
constants.  Knowledge of these force constants allows one to test the
range of interaction potential.  For that, similarly to procedure used
in \cite{sluiter}, we have calculated the absolute values of the
eigenvalues of the $3\times 3$ $\Phi _{ij}({\bf n}, \nu ; {\bf m}, \mu
)$ matrices and plotted them against a distance between atoms $({\bf
  n}, \nu)$ and $({\bf m}, \mu)$.  The result for \tbtbt supercell is
depicted in Fig.~\ref{fig:force_const}.  It shows that the
interaction drops down at least by two orders of magnitudes within the
size of the supercell.  The scatter of points is too large to assign
any rule which could govern the decaying of the force constants
parameters with distance.  From Fig.~\ref{fig:force_const} it is
clear that the \obobo supercell is too small to be used for
calculation of dispersion curves since the considered force constants
terminate at radius of $a\sqrt{3}/2 = 0.866 a$.

The knowledge of the cummulant force constants allows one to define an
approximate dynamical matrix, in which the summation over all atoms is
confined to those atoms which reside within the volume of the
supercell.  The approximate dynamical matrix becomes equal to the
conventional one at discrete wave vectors ${\bf k}_L$ given be the
equation $\exp(2\pi i{\bf k}_L\cdot {\bf L}) = 1$ \cite{parlinski},
where ${\bf L} = ({\bf L}_1, {\bf L}_2, {\bf L}_3)$ are the lattice
vectors of the supercell.  At the wave vectors ${\bf k}_L$ the phonon
frequencies $\omega ^2({\bf k}_L)$, calculated by diagonalization of
the approximate dynamical matrix, are the same as those calculated
from the exact dynamical matrix.  For \tbtbt supercell the exact wave
vectors are at $\Gamma $, X, L, W, and two other wave vector points,
namely, the midpoint between $\Gamma $ and X along $\langle 1,0,0
\rangle$ and $\langle 1,1,1 \rangle$ directions.  The advantage of the
above described direct method is that it does not impose any limit to
the range of interaction.  When the supercell size is smaller than the
range of interaction, the direct method interpolates the dispersion
curves between the exact points.

\begin{figure}[htbp]
  \leavevmode
  \includegraphics[width=\columnwidth]{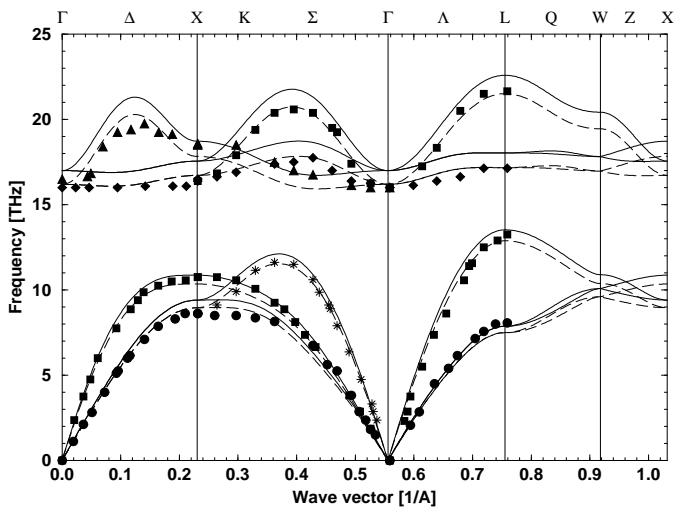}
  \caption{Phonon dispersion relations of TiC crystal 
    calculated from \tbtbt supercell (full line).  The same dispersion
    curves divided by a factor $1.05$ (dashed line). Experimental
    points taken from Ref.~\cite{pint}.}
    \label{fig:big_cell}
\end{figure}

The dispersion curves calculated on the basis of \tbtbt supercell are
shown in Fig.~\ref{fig:big_cell}.  They correspond to temperature
T=0. At the same Figure~\ref{fig:big_cell} we compare the calculated
dispersion curves with the experimental phonon frequencies measured by
the inelastic neutron scattering \cite{pint}.  Different symbols of
experimental points correspond to different phonon polarizations
established experimentally.

The experimental points are about $\Delta\omega = 0.05 \omega$ lower
then the calculated values. One of the reasons for such situation
could be that experimental values are measured at room temperature and
anharmonic effects might deminish the phonon frequencies. Another
reason may be related with a non-stoichiometric concentration of
carbon in measured samples, which diminish in average phonon
frequencies as well.  The efect that experimental phonon points are at
lower frequencies then the calculated ones appears as well in other
ab-initio calculations \cite{frank,parlinski}.

The theory of lattice dynamics provides summation rules which follow
from translational and rotational invariances of the crystal. For TiC
these invariances lead to two equations, which can be used to set the
values of on-site force constant parameters. If these conditions are
satisfied the acoustic phonon branches at {\bf k}=0 would point to
$\omega=0$. Our dispersion curves, without imposed
translational-rotational invariances point at {\bf k}$=0$ to
$\omega=0.31$THz. Since the invariance conditions should be satisfied
within the {\em ab initio} and direct methods, we have not corrected
our dispersion curves displayed in Fig.~\ref{fig:big_cell} according
to this effect.

\begin{figure}[htbp]
  \leavevmode 
  \includegraphics[width=\columnwidth]{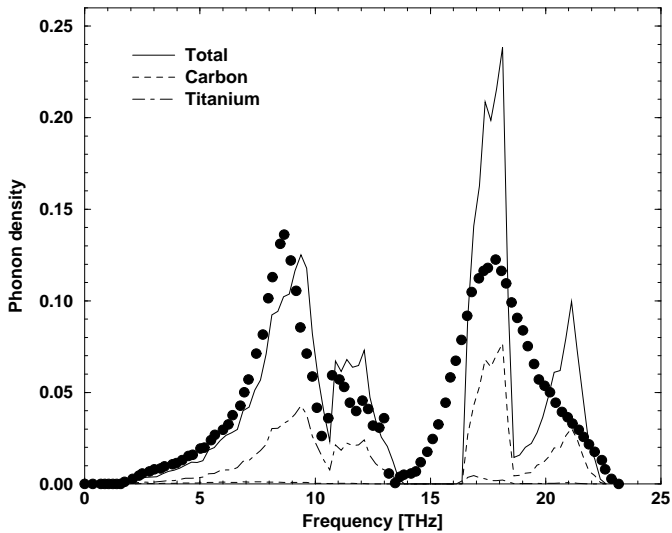} 
  \caption{Phonon density spectrum $g(\omega)$ of the TiC crystal and 
    partial phonon density spectra $g_{i,\alpha }(\omega)$,
    where $\alpha=$ C or Ti
    and $i=x,$ $y$ or $z$ are the displacement
    directions.
    Experimental points taken from Ref.~\cite{pint}.
    }
  \label{fig:statdens} 
\end{figure}

By sampling the dynamical matrix for many wave vectors, one can
calculate the phonon density function $g(\omega )$, and the partial
phonon density functions $g_{x, Ti}(\omega )$ and $g_{x, C}(\omega )$.
The $g(\omega )$ describes the number of phonon frequencies in an
interval around $\omega $, while $g_{x, Ti}$ and $g_{x, C}$ specify
the number of phonon frequencies in the interval around $\omega $ but
modified by the population of the $x$ displacement component of either
Ti or C atom.  The density of states functions are shown in
Fig.~\ref{fig:statdens}.  They are conventionally normalized to $\int
g(\omega ) d\omega = 1$ and $\int g_{x,\alpha }(\omega ) d\omega =
1/6$, where $\alpha =$ Ti or C.  The curves show that motions within
acoustic dispersion curves are almost entirely due to Ti atoms.  The
sum of Titanium density functions ($g_{x, Ti}(\omega ) + g_{y,
  Ti}(\omega ) + g_{z, Ti}(\omega )$) fits to the total density
function $g(\omega )$ below 14 THz.  Part of $g(\omega )$ above the
gap is mainly due to C atoms.  Thus, TiC form a rather special
crystal, in which the Ti atoms, as heavier, form a frame for elastic
motion, and the C atoms vibrate within the optical modes.  We add,
however, that the magnitude of the force constants between Ti - C and
between Ti - Ti and C - C at the same distances remains of the same
order.  Thus, TiC does not consist of two weakly bounded subsystems.
In Fig.~\ref{fig:statdens} the experimentally determined phonon
density of states \cite{pint,gompf} is also shown.  The agreement
within acoustic region is quite qood. In optic part the experimental
resolution of $\approx 2.5$THz and the nonsoichiometry of carbon in
the sample lead to broadening of the optic band.

The elastic constants can be evaluated from the slopes of acoustic
branches of phonon dispersion curves. Here, we have used the
invariance conditions to set the acoustic modes to $\omega=0$ at {\bf
  k}$=0$. For the fcc structure, the set of seven equations for three
unknown variables $c_{ij}$ has to be solved: $\rho v^2_{\alpha,L} =
c_{11}$; $\rho v^2_{\alpha,T} = c_{44}$; $\rho v^2_{\beta,L} =
c_{11}+4c_{44}+2c_{12}$; $\rho v^2_{\beta,T} = c_{11}+c_{44}-c_{12}$;
$\rho v^2_{\gamma,L} = c_{11}+2c_{44}+c_{12}$; $\rho v^2_{\gamma,T_h}
= c_{11}-c_{12}$; $\rho v^2_{\gamma,T_3} = c_{44}$, where the
velocities $v_i$ are the slopes of appropriate acoustic branches.
Here, $\alpha, \beta$ and $\gamma$ indexes describe three main
directions in the reciprocal space $\langle 0,0,1 \rangle$, $\langle
1,1,1 \rangle$ and $\langle 0,1,1 \rangle$, respectively.  The density
of TiC $\rho=4849.6$ kg/m$^3$ was found using the calculated lattice
constant. The solution of this set of equations, by means of the
least-squares fit, results in $c_{11}=5.7$, $c_{12}=2.5$ and
$c_{44}=1.51$ MBar.  Comparing these values with experimental data
(Table~\ref{tab:elastic}) one notices considerable discrepancies,
especially for $c_{12}$.  Closer look to dispersion curves,
Fig.~\ref{fig:big_cell}, assures us that the transverse acoustic mode
in $\langle 1,1,0 \rangle$ direction and the longitudinal in $\langle
1,1,1 \rangle$ direction do not follow exactly the experimental
points.  We have checked, that such discrepancy of slope is sufficient
to produce the observed deviations of the calculated elastic
constants, specially of $c_{12}$.  The discrepancy in slopes is so
small that it could be assigned to the missing force constants
describing mainly the long range Coulomb interaction.  Indeed, the
direct method did not handle well interactions exceeding the size of
the supercell, except for the special points $\Gamma$, X, L, W
mentioned earlier.

\begin{table}[htbp]
  \begin{tabular}{|l|D{.}{.}{4}|d|d|d|}
    \hline
    Result & 
    \multicolumn{1}{c|}{$B$} & 
    \multicolumn{1}{c|}{$c_{11}$} & 
    \multicolumn{1}{c|}{$c_{12}$} & 
    \multicolumn{1}{c|}{$c_{44}$} \\
    \hline
    Exp. TiC$_{0.91}$ \cite{exp_dataA}
    & (2.422) & 5.145 & 1.060 & 1.788 \\
    Exp. TiC \cite{exp_dataB}
    & 2.4 & 5.00 & 1.13 & 1.75 \\
    Exp. TiC$_{0.95}$ \cite{pint}
    & (2.5) & 5.4 & 1.1 & 1.8  \\
    Exp. TiC$_{0.89}$ \cite{pint}
    & (2.5) & 5.2 & 1.1 & 1.8 \\
    Calculations \cite{TiC_GGA} & 2.2 & 4.7 & 0.97 & 1.67 \\
    Ours, acoustic modes & (3.57) & 5.7 & 2.5 & 1.51 \\
    Ours, deform. energy & 2.42 & 5.39 & 0.94 & 1.92 \\
    Ours, stress--strain & (2.46) & 5.21 & 1.09 & 1.93 \\
    \hline
  \end{tabular}
  \caption{Experimental and calculated bulk modulus $B$ 
    and elastic constants $c_{ij}$
    of the TiC crystal, in units of 
    $10^{11}{\rm Nm}^{-2} = 1 {\rm MBar}$.
    Values in parenthesis are calculated from 
    $B=\frac{1}{3}(c_{11}+2c_{12})$.}
    \label{tab:elastic}
\end{table}

To rule out that the discrepancy for $c_{12}$ does not follow from
ab-initio calculations, we have derived the bulk modulus and the
elastic constants by two conventional methods \cite{TiC_GGA}. The bulk
modulus was calculated in the \obobo supercell from the total energy
as a function of the lattice constant.  This data have been fitted to
a third order polynomial.  Hence, the derivative $\frac{\partial
  E}{\partial V}$, and the bulk modulus $B$ were calculated from the
relation: $B=-V\frac{\partial E}{\partial V}$, where $V$ is the volume
of TiC crystallographic unit cell. The $B$ values are given in
Table~\ref{tab:elastic} and they agree quite well with experimental
data.

A calculation of $c_{11}$, $c_{12}$, $c_{44}$ elastic constants
involved a similar procedure, in which the crystal was deformed by
lengthening and sheer deformations.  Deformations from $0.5\%$ to
$3\%$ in length and from $1$ to $5$ degrees in angle have been used.
The results of small and large deformations are consistent.  The
deformed lattices have space groups Fm$\bar3$m, I4/mmm and I/mmm, for
bulk expansion, elongation along z, and sheer modes.  In all these
deformed lattices all atoms remain in the high-symmetry sites which
guaranties an extremum of potential energy at those positions.  Thus,
in the case of cubic TiC, one is entitled not to carry on the
relaxation of internal degrees of freedom during supercell
deformations.  The calculated elastic constants are compared to
experimental data in Table~\ref{tab:elastic}. They agree with
calculated values given in Ref.\cite{TiC_GGA} and are in good
agreement with experiment.

We have also checked the above results using stresses calculated by
CASTEP. The elastic constants were found as first derivatives of the
stress--strain relationship. For that, we have used the same runs of
electronic calculations as for the elastic constant derivation given
above using deformation dependence of energy. The results are included
in Table~\ref{tab:elastic} and they show slightly better agreement
with the experimental data.

In summary, we have shown that the {\em ab initio} calculations of HF
forces, together with the direct method lead to satisfactory
description of phonons in the TiC crystal. As usual the frequency of
the calculated dispersion curves are slightly too high.  The acoustic
modes are not sufficiently accurate, thus the elastic constants
calculated from them differ from those found through the
supercell deformations.  The acoustic part of density of states fits
well to neutron experimental data.

\begin{acknowledgement}
The computational advices of J. Oleszkiewicz and the use of
facilities of the Computer and Network Center of the
Jagellonian University, Cracow, where the calculations have
been done, are kindly acknowledged.
This work was partially supported by the
Polish State Committee of Scientific
Research (KBN), grant
No 2 PO3B 145 10.
\end{acknowledgement}



\begin{thebibliography}{99}
\bibitem{castep} CASTEP is a part of $Cerius^2$ system developed
        by Molecular Simulations.
\end{thebibliography}
\end{document}